%% file: main.tex
\PassOptionsToPackage{dvipsnames}{xcolor}
\documentclass[10pt,twocolumn,letterpaper]{article}

\usepackage{cvpr}              

\input{preamble}

\definecolor{cvprblue}{rgb}{0.21,0.49,0.74}
\usepackage[pagebackref,breaklinks,colorlinks,citecolor=cvprblue]{hyperref}

\newcommand{\twoline}[2]{\begin{tabular}{@{}c@{}}#1\\#2\end{tabular}}

\def\paperID{7234} 
\def\confName{CVPR}
\def\confYear{2024}

\title{Quantum Denoising Diffusion Models}

\author{
Michael Kölle, Gerhard Stenzel, Jonas Stein, Sebastian Zielinski,\\ Björn Ommer, Claudia Linnhoff-Popien\\
LMU Munich\\
{\tt\small michael.koelle@ifi.lmu.de}
}

\begin{document}
\maketitle

\input{content/0_abstract}    
\input{content/1_introduction}

\input{content/2_related_work}
\input{content/3_approach}
\input{content/4_experimental_setup}
\input{content/5_results}
\input{content/6_conclusion}
\section*{Acknowledgements}
This paper was partially funded by the German Federal Ministry of Education and Research through the funding program ``quantum technologies --- from basic research to market'' (contract number: 13N16196).

{
    \small
    \bibliographystyle{ieeenat_fullname}
    \bibliography{main}
}

\input{content/X_supplements}

\end{document}

%% file: preamble.tex
\usepackage[dvipsnames]{xcolor}

\usepackage{tikz}
\usetikzlibrary{quantikz2}
\usepackage{tabularx}
\usepackage{braket}

%% file: content/0_abstract.tex
\begin{abstract}
In recent years, machine learning models like DALL-E, Craiyon, and Stable Diffusion have gained significant attention for their ability to generate high-resolution images from concise descriptions. Concurrently, quantum computing is showing promising advances, especially with quantum machine learning which capitalizes on quantum mechanics to meet the increasing computational requirements of traditional machine learning algorithms. This paper explores the integration of quantum machine learning and variational quantum circuits to augment the efficacy of diffusion-based image generation models. Specifically, we address two challenges of classical diffusion models: their low sampling speed and the extensive parameter requirements. We introduce two quantum diffusion models and benchmark their capabilities against their classical counterparts using MNIST digits, Fashion MNIST, and CIFAR-10. Our models surpass the classical models with similar parameter counts in terms of performance metrics FID, SSIM, and PSNR. Moreover, we introduce a consistency model unitary single sampling architecture that combines the diffusion procedure into a single step, enabling a fast one-step image generation.
\end{abstract}

%% file: content/1_introduction.tex
\section{Introduction} \label{sec:introduction}

\begin{figure}[tb]
    \centering
    \includegraphics[width=\linewidth]{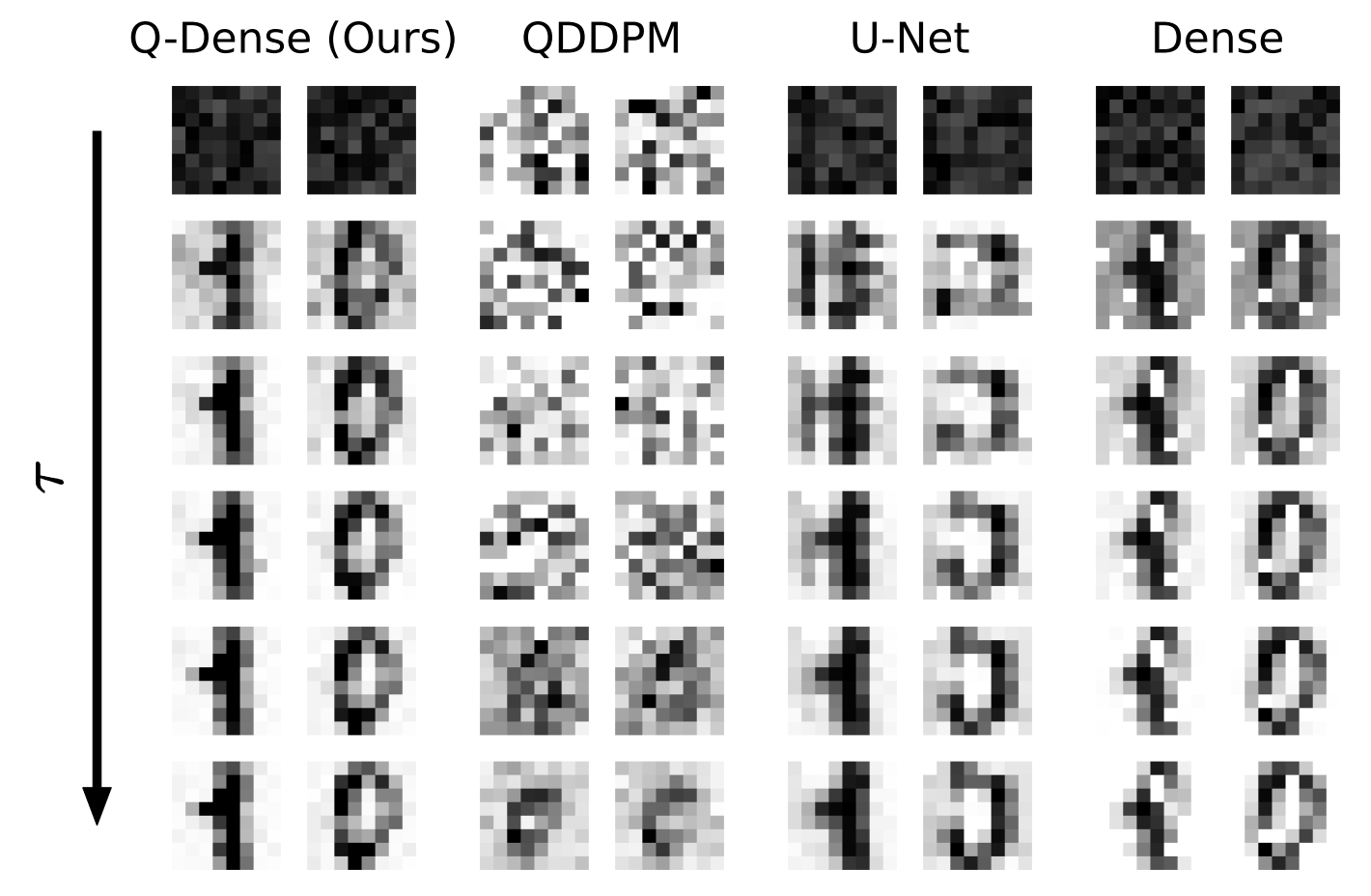}\\
    \caption[Samples on MNIST]{Diffusion process ($\tau = 10$) of models Q-Dense, QDDPM \cite{qddpm}, U-Net and Dense on MNIST digits. Samples from every second step are depicted.
    }\label{fig:mainsamples}
\end{figure}

Image generation remains a vital topic in computer vision and graphics~\cite{imagegeneration1,imagegeneration2}, encompassing tasks from synthetic data creation to artistic endeavors. Generative models, like Stable Diffusion, have found applications ranging from image editing to aiding multi-modal models like GPT-4 in providing visual responses to human queries~\cite{gpt4,stablediffusion}. 
While Denoising Diffusion Models (DDMs) have recently seen significant progress, they face notable challenges such as high computational demands and the necessity for extensive parameter tuning~\cite{diffusion_beats_gan,ddpm,stablediffusion}. Recent advancements in quantum computing present opportunities to alleviate some of these challenges~\cite{qml1,qcaccel}. Specifically, quantum machine learning (QML) uses quantum principles to enhance efficiency for classical machine learning tasks~\cite{qml1,qml2,qml3}.

In this paper, we combine QML with DDMs to form quantum denoising diffusion models (QDDMs). This synthesis retains the image generation effectiveness of DDMs while benefiting from the efficiencies of quantum computing. By merging these two powerful domains, we push the frontiers of what is currently achievable in image generation, setting new benchmarks for quality and efficiency. 
We introduce a novel quantum U-Net design, employing quantum convolutions to further refine image quality. Additionally, we leverage the inherent unitary properties of quantum circuits to optimize QDDMs' sampling time, introducing our unitary single-sample consistency model architecture. We evaluate our models together with classical deep convolutional networks and U-Nets on the datasets MNIST digits, Fashion MNIST and CIFAR10 using FID, SSIM, and PSNR performance metrics. Furthermore, we showcase the single-shot image generation capabilities on simulator and on real IBMQ hardware. 
Our results show that QDDMs hold a competitive edge over classical DDMs in producing high-quality images with fewer parameters. Finally, we present a detailed empirical analysis, on the strengths and limitations of QDDMs, setting the stage for potential future explorations in this promising intersection of quantum and machine learning. In summary, our key contributions include:
\begin{itemize}
\item The inception of two novel quantum diffusion architectures: Q-Dense and QU-Net.
\item The introduction the unitary single-sample consistency model architecture.
\end{itemize}

%% file: content/2_related_work.tex
\section{Related Work} \label{sec:related-work}

\subsection{Diffusion Models}\label{sec:diffusionmodels}
Diffusion models, as first introduced by~\cite{dpm_sohl}, present a unique approach to training generative models. Rather than the adversarial battle seen in GANs~\cite{gan_goodfellow}, diffusion models focus on the steady transformation of noise into meaningful data. While the inception of this technique showed promising results~\cite{diffusion_beats_gan}, subsequent improvements like Denoising Diffusion Implicit Models (DDIM) emerged~\cite{ddpm,ddpm2,ddim}. In contrast to traditional diffusion models, which sample each intermediary step in a Markov chain fashion, DDIM identifies and removes noise earlier, bypassing certain sampling iterations~\cite{ddim}. In this work, we primarily follow the methodology detailed in Ho \textit{et al.}~\cite{ddpm}.

\subsection{Variational Quantum Circuits} \label{sec:VQC}
Quantum machine learning (QML) aims to harness the capabilities of quantum computing to meet the increasing computational requirements of traditional machine learning algorithms~\cite{qml1, qcaccel}. Variational quantum circuits (VQC) are foundational to QML, serving as function approximators similar to classical neural networks. These circuits utilize parameterized unitary quantum gates on qubits~\cite{q_ops}, leveraging the principles of quantum mechanics such as superposition, entanglement, and interference. These gates derive their parameters from rotation angles, which are trainable via conventional machine learning methods. A VQC's architecture consists of three components.

The first component embeds image and guidance data into qubits. For image data, we employ amplitude embedding, this method encodes $2^n$ features (pixel values) into $n$ qubits, representing each feature as a normalized amplitude of the quantum state~\cite{mottonen, qc_steane}. For label embedding, we use angle embedding, which only encodes $n$ features into $n$ qubits but uses less quantum gates~\cite{PhysRevLett.129.230504}. The second component consists of multiple variational layers, similar to hidden layers in classical networks. We design our circuits with strongly entangling layers, following the approach of Schuld \textit{et al.}~\cite{ccqc}. We also apply data re-uploading, re-embedding parts of the input in-between variational layers, which aids in more complex feature learning~\cite{datareuploading}. Lastly, we extract the output by measuring the quantum system, causing the system's superposition to collapse. Given \(n\) qubits, this allows us to derive \(2^n\) joint probabilities of the output states. Quantum simulators further enable the extraction of the circuit's state vector, which we use to build the combined unitary matrix in \cref{sec:methods:singlepasstheory}.

It's important to highlight that while VQCs can efficiently manage high-dimensional input with just $\log_2(N)$ qubits~\cite{qml2}, they still encounter issues such as high qubit costs and error rates in the current Noisy Intermediate-Scale Quantum (NISQ) era~\cite{nisq}. However, anticipated advancements in these domains hold promise for QML's pivotal role ahead~\cite{nisq,qcaccel}.

\subsection{Quantum Diffusion Models}\label{sec:rel:qddpm}

To the best of our knowledge, the model QDDPM by Dohun Kim \textit{et al.} currently stands as the sole quantum diffusion method for image generation~\cite{qddpm}. They designed a single-circuit model with timestep-wise layers that take unique parameters for each iteration, and shared layers consistent across all iterations. This model shines in its space-efficiency, needing only $\log_2(\text{pixels})$ qubits and thus exhibiting logarithmic space complexity for image generation. To counteract the vanishing gradient issue, they constrained the circuit depth. For entanglement, they utilized special unitary (SU) gates, targeting two qubits simultaneously. While "SU(4)" groups offer benefits like known differentiation, their parameter efficiency per gate is lacking, as they are using 15 parameters per group. Given the constrained circuit depth, their model produces images that are somewhat recognizable but miss the intricacies of the originals (refer to \cref{fig:mainsamples}).  

%% file: content/3_approach.tex
\section{Quantum Denoising Diffusion Models} \label{sec:approach}

\subsection{Dense Quantum Circuits}\label{sec:dense_quantum_circuits}

In our work, we employed a dense quantum circuit (or strongly entangling circuit) as the foundational component of our quantum models. The term ``dense'' refers to the extensive entanglement among qubits in the circuit. This design choice is reminiscent of the nomenclature in classical deep learning, where the term ``dense'' or ``fully connected'' describes layers where every neuron is connected to every other neuron in adjacent layers.

The architecture of our dense quantum model is a follows. As detailed in \cref{sec:VQC}, we chose amplitude embedding for input embedding due to its space-efficiency. Given that we are training on simulators, we bypassed the initial preprocessing steps outlined in~\cite{mottonen}. Instead, we directly initialized the normalized data to the state of the quantum circuit. We encode the normalized class indices for guidance by utilizing angle embedding. This is achieved by adding an additional qubit (ancilla) and performing a rotation around the x-axis by an angle of \(\text{class index} \times \frac{2\pi}{\text{\#classes}}\). Our circuit's variational component consist of several strongly entangling layers~\cite{ccqc}, resulting in a a total of $ \text{\#layers} \times 3 \times \text{\#qubits}$ trainable parameters. We then calculate the joint probabilities of a qubit subset, measuring the likelihood of the output being in states $\ket{00\dots00}$ to $\ket{11\dots11}$. If our output vector surpasses the input vector in size, we truncate the excess, eliminating unused measured probabilities. To align the output within the input data's range, we scale the obtained probabilities using the input data's euclidean norm. 

\subsection{Quantum U-Net}\label{sec:qunet}

\begin{figure}[htb]
    \centering
    \includegraphics[width=\linewidth]{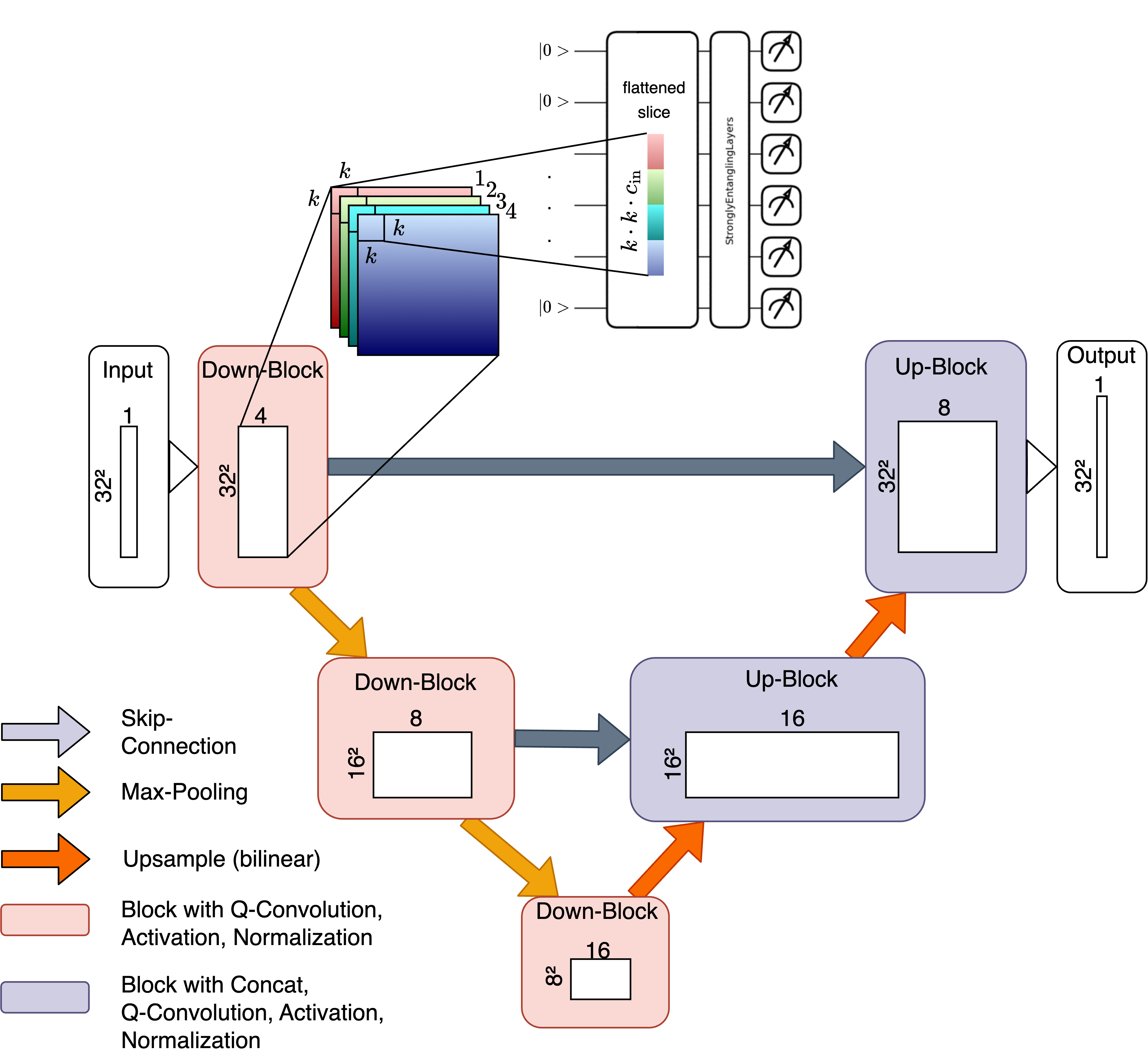}\\
    \caption[QU-Net architecture]{
        QU-Net architecture and quantum convolution, embedding a flattened slice into a dense quantum circuit.
    }\label{fig:unetstructure2}
\end{figure}

Our Quantum U-Net (QU-Net), draws inspiration from classical U-Nets, particularly those without attention layers and upscaling features (shown in \cref{fig:unetstructure2}). 

Opposed to the well-known blocks with two or more classical convolutions each, we incorporate only one quantum convolution layer per block, as we observed prolonged execution times when using more convolutions. Quantum convolutions are our novel approach to use the flexibility of convolutions in quantum machine learning, allowing us to embed any slice of shape $c_\text{in}\times k\times k$ into a dense quantum circuit (\cref{sec:dense_quantum_circuits}) with $\max(\log_2(c_\text{in}\times k\times k), \log_2(c_{\text{out}}))$ wires (\cref{fig:unetstructure2}) and measure $c_{\text{out}}$ outputs (with $c$ being input and output channels, and $k$ being the kernel size), thus differing from existing solutions like Quanvolution \cite{qconv1} and Quantum CNNs \cite{qconv2,qconv3}.

\subsection{Guidance}\label{sec:methods:guidance}

Diffusion models can be extended with guidance, introducing auxiliary data during both training and inference. This process, represented as \(p_\theta(x_{t-1}|x_{t}, c)\) (or \(p_\theta(\epsilon_{t-1}|x_{t}, c)\)), uses \(c\) as the guiding data~\cite{glide}.

For our dense quantum circuit, normalized class labels are embedded as rotation angles into an additional ancilla qubit, ensuring distinct quantum state representation. For instance, labels 0 and 1 correspond to angles 0 and \(\pi\). Classical dense networks traditionally introduce inputs via an extra neuron per layer, enhancing performance and increasing parameter count. 

Contrarily, U-Nets, due to their architecture, implement a mask encoding for labels. This mask, defined as \(mask(c)= 0.1 \cdot \sin(c+\text{height}/20)\), subtly alters the input image with strategically placed pixel value stripes, facilitating label identification. For quantum U-Nets, despite the necessity for normalized inputs in quantum convolutions, this masking technique remains effective. However, datasets with extensive class variety might demand alternative strategies.

\subsection{Unitary Single Sampling}\label{sec:methods:singlepasstheory}
\begin{figure}[htb]
    \centering
    \includegraphics[width=\linewidth]{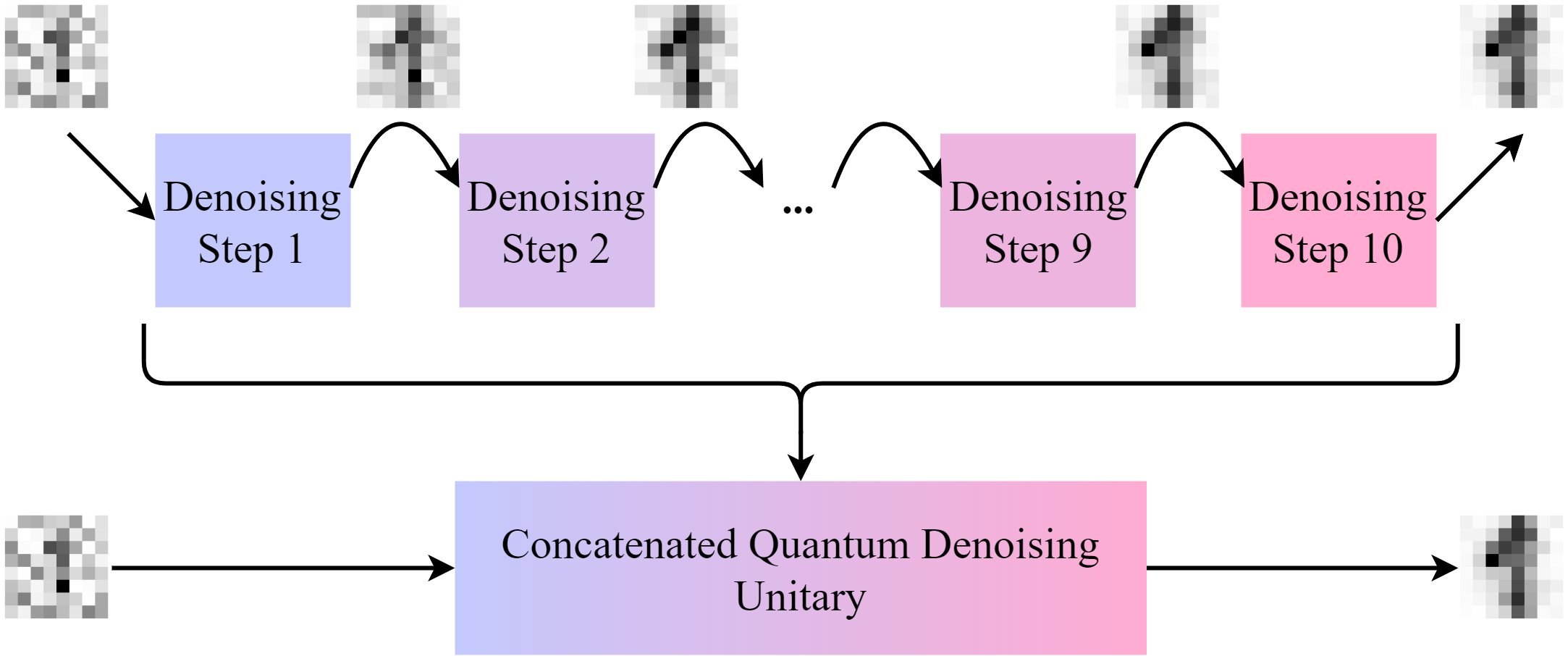}\\
    \caption[Structure of USS]{
        Unitary Single Sampling architecture.
    }\label{fig:singlesample}
\end{figure}

The unitary nature of quantum gates and circuits allows us to combine the iterative application of $U^{\tau}$ during a diffusion step into one unitary matrix $U$ (\cref{fig:singlesample}). This enables us to create synthetic images, using a single-shot of the circuit $U$, bridging the gap between quantum diffusion models and classic consistency models~\cite{consistency}. Additionally, this approach can be faster than executing multiple iterations of a classical diffusion model or even faster than executing each gate individually, depending on transpliation process and quantum hardware.

Training the single sampling model demands an alternate loss computation. Instead of typical measurement probabilities, we are interested in the post-circuit quantum state. Training directly on quantum computers is currently infeasible as quantum state tomography for state reconstruction scales exponentially with system size~\cite{cramer2010efficient}. Therefore, we utilize a noise-free quantum simulator. Loss gets evaluated by comparing the post-circuit state $p_{\theta}( \widehat{ \ket{x_{t-1}}} | \ket{x_t} )$ and the less noisy image $\ket{x_t}$, both represented as $2^n$-length complex vectors, using metrics like Mean Absolute Error (MAE).

For efficient sampling, we employ the trained parameters in the concatenated circuit $U^{n}$. Alternatively, we can determine a singular $U_\text{diffusion}$ matrix from the matrix form of $U^{n}$. Precomputing this matrix allows for more streamlined sampling~\cite{consistency}.




%% file: content/4_experimental_setup.tex
\section{Experimental Setup}\label{sec:expsetup}

\subsection{Datasets}\label{sec:datasets}
We use three well-known datasets to evaluate and compare our models: the MNIST digits~\cite{mnist}, Fashion MNIST~\cite{fashionmnist}, and a grayscale version of CIFAR-10~\cite{cifar}.

The MNIST dataset features $28 \times 28$ pixel grayscale images of digits (0-9). To probe scalability, we utilize the original as well as downscaled ($8 \times 8$) and upscaled ($32 \times 32$) versions. For specific experiments tailored to quantum circuit embedding, we narrow our focus to digits 0 and 1.
Fashion MNIST, in contrast, presents a multitude of intra-class and inter-class variations. A pivotal challenge is that models must interpret extensive regions of an image, beyond just the central figures, as peripheral areas do not consistently register as zero.
The CIFAR10 dataset stands out due to its varied backgrounds and the challenges brought forth by reduced edge contrast in grayscale. We've adopted a grayscale version created by averaging the RGB channels of its $32 \times 32$ pixel images. 

\subsection{Metrics}\label{sec:quality_measures}
To gauge the quality of our generated images, we employ three metrics: the Fréchet Inception Distance (FID)~\cite{fid1,fid2}, the Structural Similarity Index Measure (SSIM)~\cite{ssim}, and the Peak Signal-to-Noise Ratio (PSNR).

The FID serves as a tool to gauge the resemblance between original and generated data. It achieves this by calculating the Wasserstein-2 distance between Gaussian distributions of activations derived from the Inception-v3 model~\cite{inceptionv3}. A noteworthy aspect of FID is that lower scores suggest a closer resemblance between datasets~\cite{fid1}.
SSIM is defined as
\begin{equation}
    \label{eq:ssim}
    \text{SSIM}(x,y) = \frac{ 2 \mu_x \mu_y + c_1 }{ \mu_x^2 + \mu_y^2 + c_1 } \cdot \frac{ 2 \sigma_{xy} + c_2 }{ \sigma_x^2 + \sigma_y^2 + c_2 },
\end{equation}
and offers a measure of similarity between images $x$ and $y$. Higher SSIM scores are indicative of more significant image resemblance.
Lastly, the PSNR stands as a metric to quantify the noise levels within an image. Superior image quality is represented by higher PSNR values.

\subsection{Baselines} \label{sec:baselines}
Our benchmarking process contrasts our models against the architectures Deep Convolutional Networks (DCNs), U-Nets, and Quantum Denoising Diffusion Probabilistic Models (QDDPM).

DCNs are hierarchical models that utilize convolutional layers to extract progressively complex spatial features from input data \cite{cnn1,cnn2,cnn3}. By leveraging spatial invariance through weight sharing and pooling operations, DCNs can discern intricate patterns in large-dimensional datasets. Their depth and specialized architectures, such as residual and inception modules, enable the capture of both low-level image details and high-level semantic information, making them integral to advanced computer vision tasks.

The U-Net architecture~\cite{unet,unet2,unet3} is acclaimed for its capabilities in image segmentation. The U-Net's encoder captures local features, while its decoder combines these insights with broader context through skip connections, promoting gradient flow and information transfer~\cite{skip,skip2}. Strategies such as zero-padding and interpolation techniques ensure image sizes remain consistent across the iterative diffusion model process~\cite{ddpm2,diffusion_beats_gan}. Further refining image quality, attention layers become particularly beneficial when integrated with natural language embeddings~\cite{stablediffusion,attention}.

Lastly, we compare to the quantum state-of-the-art QDDPM by Dohun Kim \textit{et al.}~\cite{qddpm} as described in \cref{sec:rel:qddpm}. This model is specially designed for the $8\times8$ MNIST dataset, incorporating six gates per layer, which results in a total of 990 parameters over $\tau=10$ timesteps. When adapted for $16\times16$ images, the model requires a greater number of parameters, totaling 4920. It is important to note that our comparisons are qualitative in nature, as we lack common metrics for evaluation.

\subsection{Model Training and Evaluation}
We build our quantum models using the PennyLane framework~\cite{pennylane}. For training our quantum models, we use the PyTorch integration of PennyLane which facilitates classical backpropagation for the gradients w.r.t. the rotation angles. On actual quantum hardware, parameter-shift differentiation calculates gradients by re-evaluating circuits with perturbed parameters~\cite{acl_mitarai,qgrad}. To enhance convergence and stabilize training, parameter remapping confines values within the range $[-\pi,\pi]$ or $[0,2\pi]$~\cite{weightremapping}. We used the classical optimization algorithm Adam~\cite{adam} and minimize the Mean Squared Error (MSE) between the generated image \(p_\theta(x_{t})=\widehat{x_{t-1}}\) and \(x_{t-1}\), sourced from a noise-augmented training dataset. Notably, for the unitary single-sample model, we adopt the mean absolute error (MAE) due to its native PyTorch implementation for complex tensors. All runs were conducted on identical hardware with Intel Core \textsuperscript{®} i9-9900 CPUs and 64 GB of RAM.
In our two preliminary studies (\cref{sec:preliminary_studies}), we explore the relationship between model hyperparameters and metrics. Additionally, we performed a hyperparameter search focusing on learning rate and batch size. The detailed hyperparameter settings for each model are available in \cref{sec:hyperparameters}.
Regarding our inpainting task experiments~\cite{inpainting,inpainting2}, we evaluate the models without specific training for this purpose, using MSE to assess image fidelity. Challenges emerged when masks hid essential features: unguided models predominantly depended on existing pixels, whereas guided models benefitted from label guidance.

%% file: content/5_results.tex
\section{Experiments} \label{sec:results}
In our experiments, we assess the effectiveness and efficiency of our quantum models across various datasets and conditions. We utilize diverse datasets, including MNIST Digits \(8\times8\), Fashion MNIST \(28\times28\), and CIFAR10 \(32\times32\), to explore our models' performance under varying data complexities and dimensionalities. Our novel Unitary Single Sampling approaches are tested in several scenarios: MNIST Digits \(8\times8\) both unguided without ancilla and guided with ancilla, MNIST Digits \(32\times32\) unguided without ancilla, and MNIST Digits \(8\times8\) unguided without ancilla on IBM Q hardware. Detailed results of these experiments, highlighting key findings and observations, are presented in the following sections.

\subsection{MNIST Digits}\label{sec:res:comparison:smallguided}
We analyze performance by creating models with varying layer sizes for each dataset and measure their complexity via the sum of trainable parameters. For guided MNIST $8\times8$ images, models encompassed approximately 1000 parameters. All trained models with their respective configuration can be found in \cref{tab:lr-nq} and \cref{tab:lr-q}.

Quantum-wise, we employed a dense circuit (Q-Dense) with 47 layers and 7 qubits, utilizing 6 qubits for image embedding and measurement, plus an additional one for label embedding. A preliminary comparison among quantum models revealed small advantages via data re-uploading ~\cite{datareuploading}, but introduced a challenge due to embedding gates altering the circuit's quantum state. The best model (7 re-uploads, red-line) scored around 10 FID points lower than the no re-upload model, although the difference dwindled with an increasing number of re-uploads.

For our comparison, we use fully-connected classical networks with 1000 trainable parameters and a U-Net of total depth 2, having 3 channels in the first block and 6 in the second. We compared our models to U-Nets with a depth of 3, and 2 or 4 channels in the initial block, even though they exceeded the 1000 parameter limit. 

\begin{figure}[htb]
    \centering
    \includegraphics[width=\linewidth]{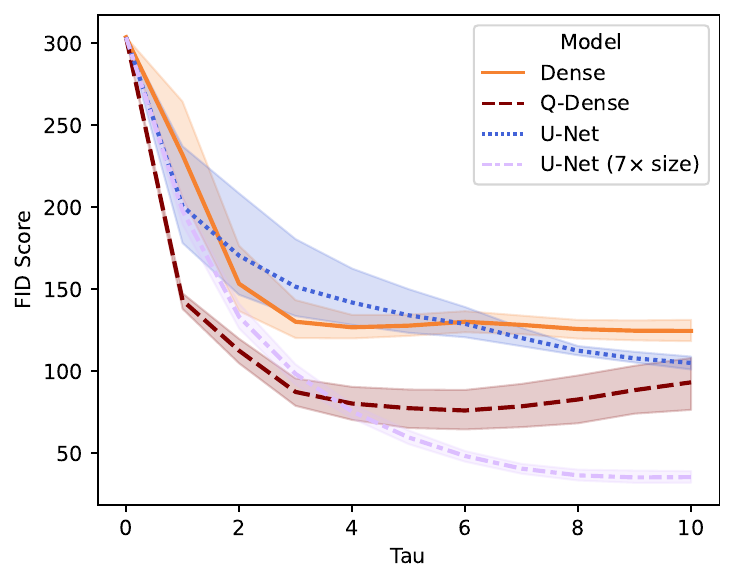}
    \caption[FID scores on MNIST 8x8 guided]{FID scores on MNIST 8x8 with guided models. $\tau$ denotes the diffusion steps. Lavender line illustrates larger U-Net capabilities for reference.
    }\label{fig:res:comparison:smalguided:fid}
\end{figure}

The Q-Dense model significantly surpassed its classical equivalents, having the same number of parameters, and showed exceptional performance especially when $\tau$ values were in the range of $3$ to $5$. This is particularly remarkable considering that all models were trained with $\tau=10$, highlighting the Q-Dense models' advanced ability to learn from the original data distribution. However, a drawback is observed when excessive iterations are performed, leading to a decline in FID score caused by ongoing modifications to the input image, which ultimately produces artifacts.

The purple line represents the largest U-Net model (with channels sized 4, 8, and 16). This model outperformed all other architectures and has more than seven times the number of trainable parameters. Quantum models outperformed classical models with similar parameter counts, achieving FID scores around 100, thus 20 points better than classical models. The quantum models, along with the largest U-Net, exhibited slightly more consistent lower score-variance than other classical models across all runs, indicating more consolidated knowledge.

\subsubsection{Inpainting}\label{sec:res:comparison:small:inpainting}

\begin{figure}[htb]
    \centering
    \includegraphics[width=\linewidth]{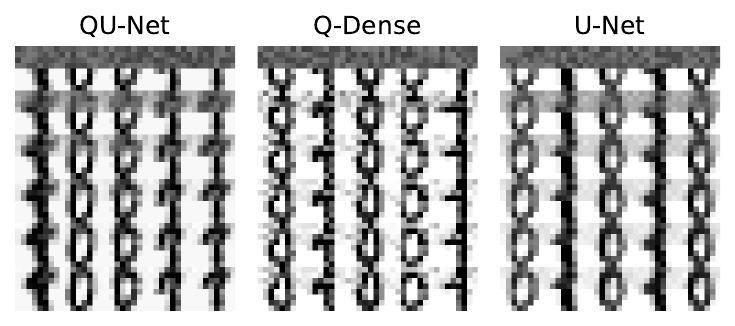}
    \caption[Inpainting samples (small bottom mask, with reset)]{
        Inpainting samples with a small mask on the top half, resetting the bottom after each of the 10 iterations.
    }
    \label{fig:res:comparison:inpainting:samples:smallmaskwreset}
\end{figure}

We used MSE to evaluate the inpainting capabilities of models with $\approx1000$ parameters, testing various masks and noise conditions across multiple scenarios, illustrated in \cref{fig:res:comparison:inpainting:samples:smallmaskwreset}. Notably, the dense quantum circuit produced visually consistent samples with minor artifacts while maintaining high overall quality. Despite presenting a better FID score, the deeper quantum U-Net performed worse compared to its shallower counterpart. 

In a experiment, where the original pixel get reset after the inpainting, most models showed declining performance after initial steps, with only the deep convolutional network maintaining consistent \cref{fig:res:comparison:inpainting:samples:smallmaskwreset}, albeit low-quality, output. Sample quality consistently held across varied masks and predictive models. In conclusion, our quantum models successfully performed knowledge-transfer tasks without specific inpainting training. They achieved satisfactory inpainting results and MSE scores, which were only marginally lower than those of classical networks, despite the classical networks having twice as many parameters.

\subsection{Fashion MNIST}\label{sec:res:comparison:fashionguidednoise}

We trained models with approximately 4000 parameters on a subset of the Fashion dataset, focusing on the "T-Shirt/Top" and "Trouser" classes due to their relative structural similarity and middle-ground complexity between the MNIST Digits and CIFAR dataset. This dataset, containing more outliers and variances than MNIST Digits and featuring images of $28\times28$ pixels, necessitated notably longer training times for models to stabilize.

\begin{figure}[htb]
    \centering
    \subfloat{
        \includegraphics[width=0.49\linewidth]{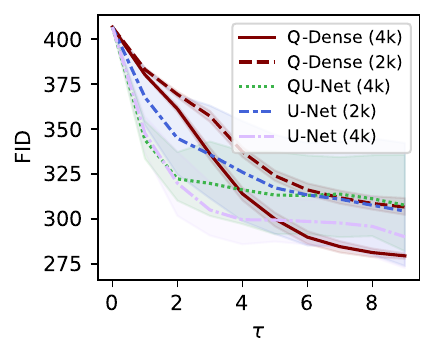}}
    \subfloat{
        \includegraphics[width=0.49\linewidth]{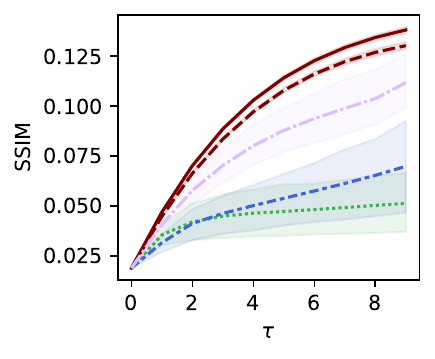}}
    \caption[Sample Quality on the Fashion dataset (FID and SSIM)]{
        Model sample quality on the Fashion dataset, evaluated using FID and SSIM.
    }
    \label{fig:res:comparison:fashion:fidssim}
\end{figure}

Upon stabilization, the models exhibited good performance, with the larger dense quantum circuit achieving the top FID score of 280, as depicted in \cref{fig:res:comparison:fashion:fidssim}. Our QU-Net and a classical U-Net achieved comparable FID scores. In terms of structural similarity, the Q-Dense models surpassed the (Q)U-Nets. 

\begin{figure}[htbp]
    \centering
    \includegraphics[width=\linewidth]{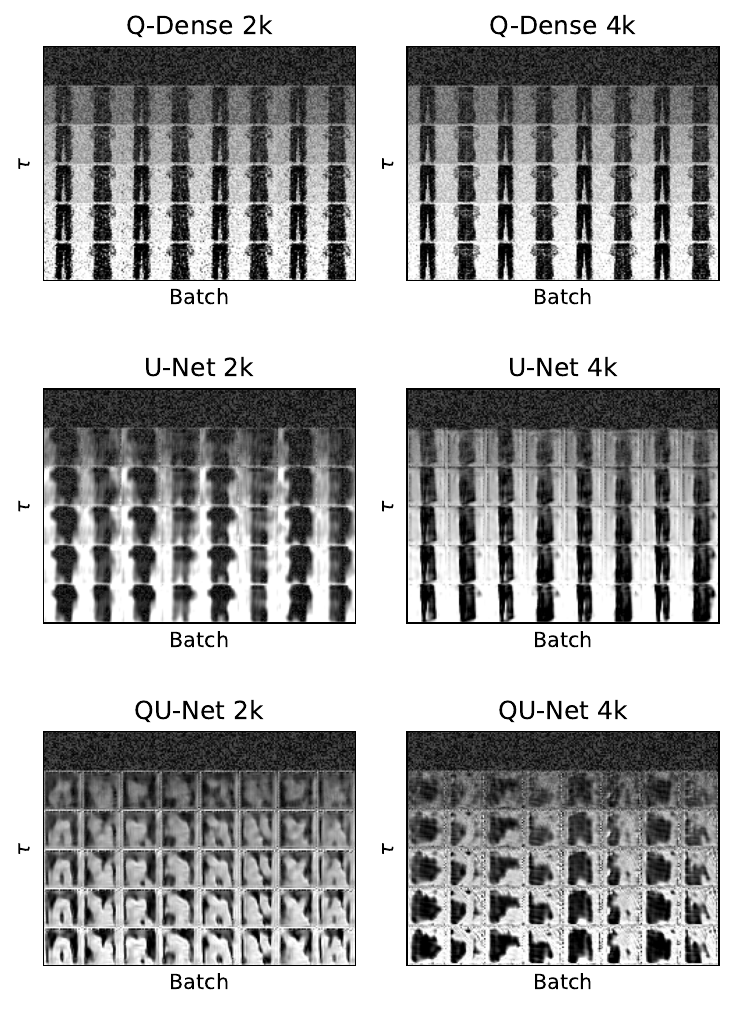}
    \caption[Samples on the Fashion dataset]{
        Chosen samples from the Fashion dataset, displaying every third $\tau$ and selected for optimal FID score per architecture.
    }
    \label{fig:res:comparison:fashion:samples}
\end{figure}

Examining the samples (\cref{fig:res:comparison:fashion:samples}), dense quantum circuits generated well-defined images with noticeable noise, while classical U-Nets produced less noisy, albeit less discernible, shapes. The quantum models achieved higher scores in SSIM, focusing on general structure, while FID, sensitive to noise, offered mixed results, thus rendering inconclusive the performance comparison between the small dense quantum circuit and the U-Net. PSNR scores mirrored SSIM results but also highlighted a lower performance in the quantum U-Nets, with the smaller QU-Net particularly affected by substantial artifacts along image edges. Discussion on potential solutions to observed issues with dense quantum circuits is available in \cref{sec:conclusion}.

\subsection{CIFAR10}\label{sec:res:comparison:cifar}

We compared models on the CIFAR10 dataset, illustrating the limitations of our quantum models: low output fidelity, potential for mode collapse, and slow execution times.
Low output fidelity arose primarily from the measurement process's mathematical properties, wherein the output was always normalized (\cref{sec:VQC}). Applying our approach of multiplying the output by the norm of the input state was only effective for homogeneous datasets like MNIST Digits. For non-homogeneous datasets like CIFAR10, this approach could result in over- or underexposed images.
We observed a heightened risk of mode collapse due to the vast difference in modes within the CIFAR10 dataset. Our models summed their losses across all batches during training, meaning that dataset outliers could distort the gradient landscape, hindering learning of the full distribution. We employed varying guided model configurations for quantum models and a $\approx 1800$ parameter U-Net, despite seemingly small for the dataset, to manage training time.

\begin{table}
    \centering
    \caption{Average metrics on CIFAR}
    \begin{tabularx}{0.8\linewidth}{Xrrr}
        \toprule
        Model        & FID              & PSNR            & SSIM            \\
        \midrule
        U-Net         & 395.47           & 8.95            & 0.026           \\
        QU-Net       & \bfseries 271.03 & 9.60            & \bfseries 0.086 \\
        Q-Dense       & 399.38           & \bfseries 10.06 & 0.061           \\
        \bottomrule
    \end{tabularx}
    \label{tbl:cifar-metrics}
\end{table}

As evidenced in \cref{tbl:cifar-metrics}, quantum U-Nets outperformed classical U-Nets with the same parameter count, and some smaller QU-Nets achieved superior FID scores. Conversely, the dense quantum models, limited by their higher-dimensional input state and therefore the number of layers to prevent memory issues, exhibited the weakest performance.


In the generated samples, small QU-Nets with 1000 parameters generated mostly large-scale structures, exhibiting a high variance. The larger QU-Net with 4000 parameters generated more detailed images with finer structures. The classical U-Net (2000 parameters) produced styles intermediate between the two quantum U-Nets. Meanwhile, the dense quantum model performed the weakest, displaying little discernible structure in the samples.

Despite the evident limitations of the dense quantum model, quantum U-Nets demonstrated superiority over classical U-Nets, affirming the advantages of our quantum convolution layers over classical convolutions. Nonetheless, they also exhibited significant drawbacks, such as notably higher training and sampling times.

Upon comparing the training and sampling speeds of models, simplified U-Nets (134 seconds per training epoch, 859 seconds per sampling run) proved much slower than Q-Dense models (2 and 9 seconds) and exhibited better scaling with the number of parameters. The U-Nets, hindered by a high parallel batch size and a sequential backward pass, were notably influenced by the high parallel complexity. We mitigated this with caching to expedite the training process, despite the requisite higher memory amount.

In conclusion, while U-Nets' sublinear scaling was not beneficial due to a bottleneck in the simulator framework, leading to exceedingly slow execution, preliminary experiments with caching the matrix representations of the quantum layers demonstrated potential for future improvement.

\section{Unitary Single-Sampling}

In this exploration of our novel unitary single-sampling models, we benchmarked them against each other in terms of sample quality, training duration, and speedup, rather than against conventional models. We selected a subset of the MNIST dataset featuring $8\times8$ and $32\times32$ images labeled 0 and 1, and trained versions with guided, unguided, and unguided with ancilla qubit approaches. Moreover, we executed some models on IBMQ's 7-qubit quantum hardware since our bit-efficient models required only $\log_2(8\times8)$ qubits.

Matrix transformations of certain trained circuits showcased the performance enhancement of condensed representation. We examined various noise initializations for inference and their resultant impact on image quality, utilizing amplitude embedding to encode random noise for inputting initial noise $x_\tau$. Single-pass models, requiring only a single forward pass during sampling and predicting data reconstruction $p_\theta(x_{t-1}|x_t)$, exhibited their advantage when treating the concatenation of repetitions as a singular circuit.

\begin{figure}[htb]
    \centering
    \subfloat[Without ancilla]{
        \includegraphics[width=0.49\linewidth]{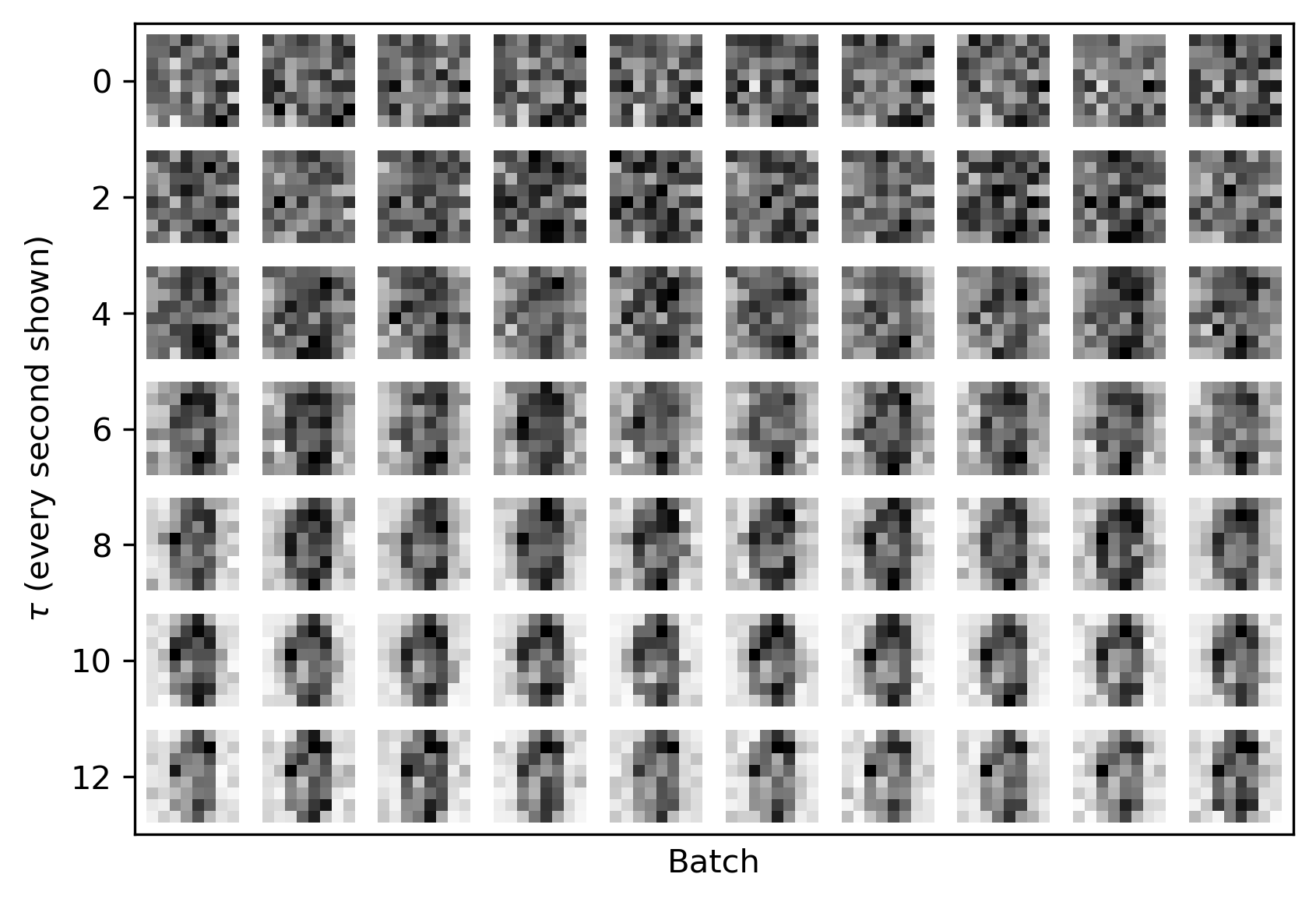}
        \label{fig:ss:8x8undirnoanc:56:sample}}
    \subfloat[With ancilla]{
        \includegraphics[width=0.49\linewidth]{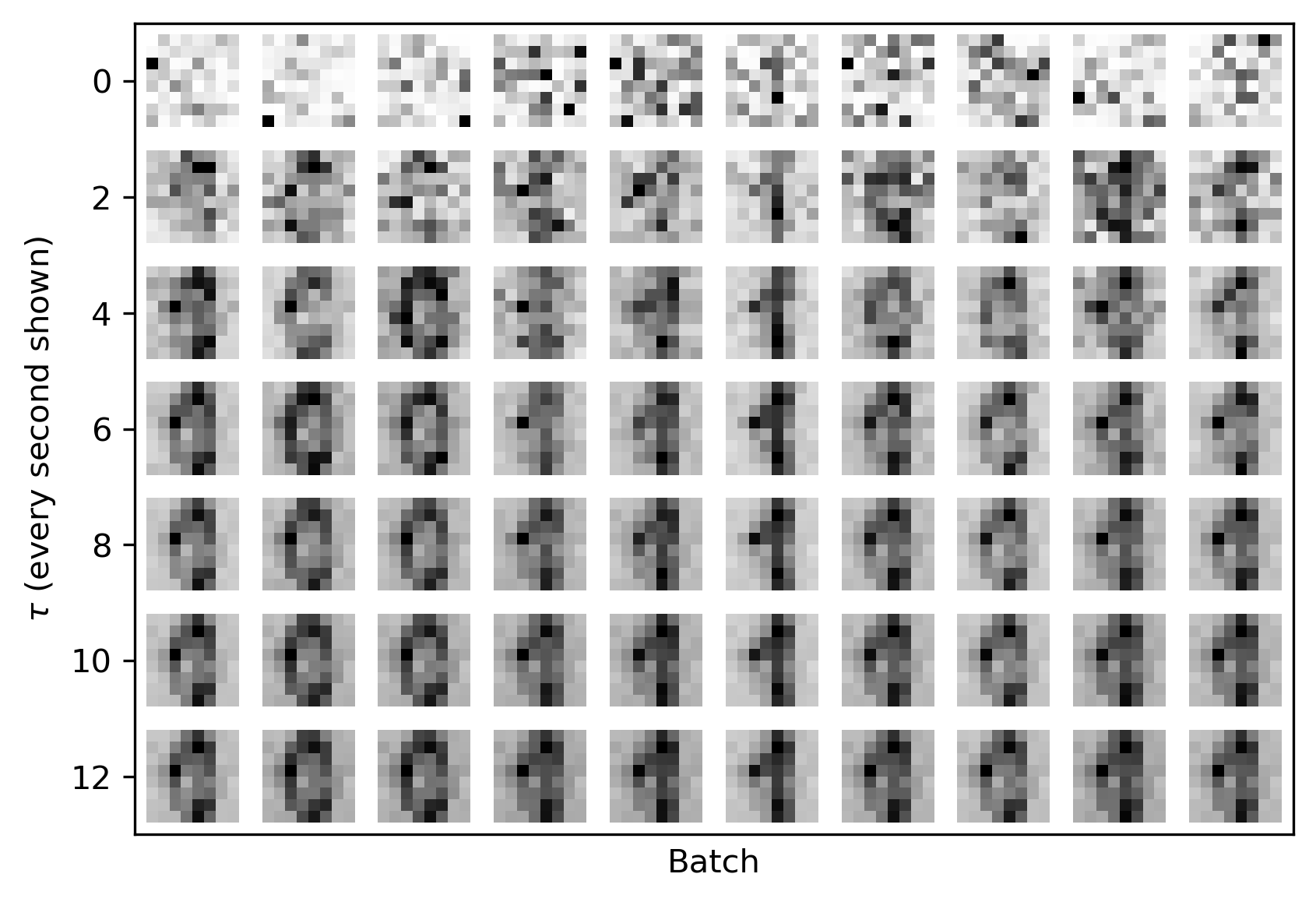}
        \label{fig:ss:8x8undiranc:47:sample}}
    \caption[Undirected single-sample small MNIST samples]{
        Samples from undirected single-sample models on the MNIST $8\times8$ dataset.
    }
    \label{fig:ss:8x8undir}
\end{figure}

\textbf{MNIST Digits $8\times8$} The 56 quantum layer, undirected model without ancilla qubits, was transformed into matrix representation, streamlining the sampling process from $1008 \tau$ quantum gate applications to a single matrix multiplication and vastly accelerating the sampling process. Conversion took under 5 seconds for $\tau=1$ and was executed once due to class-independence, while repeated multiplications for other $\tau$ values took mere microseconds in the \textit{PyTorch} framework. The model successfully generated discernible digits despite low quality, with its performance surpassing that of its 28-layer counterpart. Although models produced better samples with $8 < \tau < 14$, both were capable of generating distinguishable samples from both classes with adequate training.

\begin{figure}[htb]
    \centering
    \subfloat{
        \includegraphics[width=0.49\linewidth]{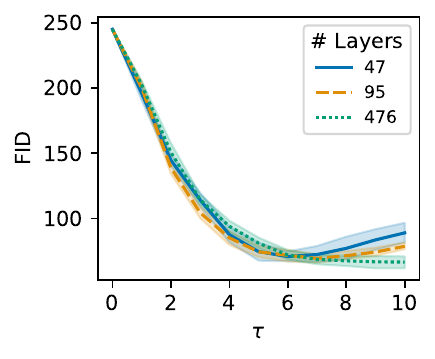}}
    \subfloat{
        \includegraphics[width=0.49\linewidth]{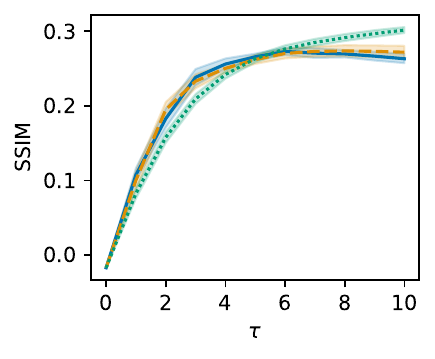}}
    \caption{
        FID and SSIM scores of guided single-sample models on the MNIST $8\times8$ dataset
    }
    \label{ffig:ss:dirfidssim}
\end{figure}

\begin{figure}[htb]
    \centering
    \subfloat[Simulator]{
        \includegraphics[width=0.78\linewidth]{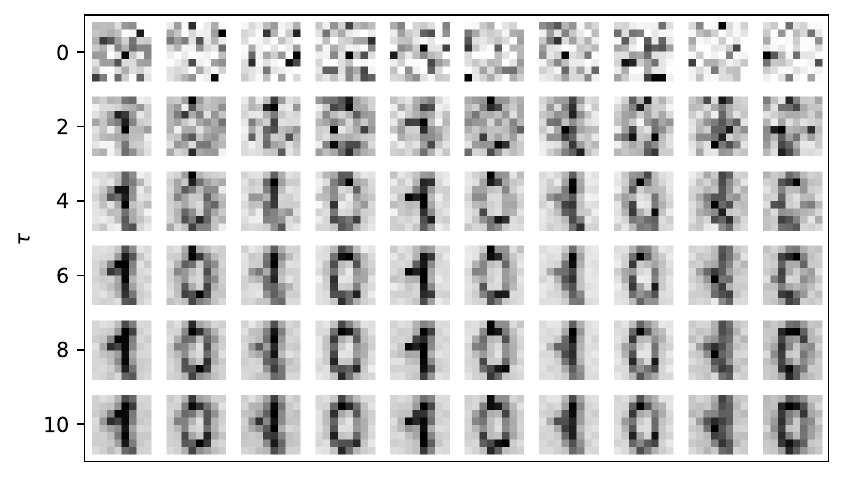}
        \label{fig:ss:dirsamples}}
    \subfloat[IBMQ]{
        \includegraphics[width=0.20\linewidth]{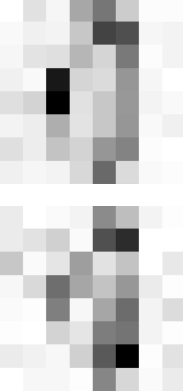}
        \label{fig:ibmq_mnist8x8_ss}}
    \caption{
     Unitary Single-Sampling Samples on Simulator (guided) and on IBMQ (unguided). 
    }
    \label{ffig:ss:samples-uss}
\end{figure}

Directed single-sample models of various architectures were trained, revealing a trend where larger models  benefited from added guidance, as evidenced by a 17\% FID score improvement and an SSIM score boost, whereas smaller models could be impeded by excessive re-uploads. Evidently, single-sample models generally profited from a larger number of trainable parameters across all metrics, as depicted in \cref{fig:ss:dirsamples}. These models successfully generated recognizable digits from a normal distribution without requiring custom multinomial distributions.\\

\textbf{MNIST Digits $32\times32$} Given the 10-qubit image representation requirement, we limited our evaluation to unguided models without ancillas due to simulator constraints. As with the $8\times8$ dataset, the 133-layer model (approx. 4000 parameters) outperformed the 66-layer model. However, both models struggled to represent the training manifold, consistently generating only zeros. Notably, the larger model produced recognizable digits up to $\tau=16$, underscoring the scalability of our approach. Despite limitations in class representation, these models retained some diffusion model properties, requiring only $\tau$ matrix multiplications for results, unlike traditional models.\\

\textbf{MNIST Digits $8\times8$ on IBMQ} We ran our unguided no-ancilla model on IBMQ's quantum hardware. As \cref{fig:ibmq_mnist8x8_ss} shows, it produced distinguishable samples. Despite inherent quantum hardware noise, with 10000 shots, our fully-quantum circuit successfully executed a diffusion model in roughly 40 seconds, excluding transpilation and queuing. Classical post-processing was minimal. Our diffusion models' inherent noise robustness proved beneficial, acting as intrinsic error correction, making them suitable for the NISQ era. As lower-noise quantum hardware emerges, our model's quality will likely improve. However, experiments on older calibrated devices yielded less recognizable digits.

%% file: content/6_conclusion.tex
\section{Conclusion} \label{sec:conclusion}
In our research, we explored quantum denoising diffusion models, introducing the Q-Dense and QU-Net architecture. Furthermore, we introduced a quantum consistency model called unitary single-sampling, which consolidates the diffusion process into one unitary matrix, enabling one-step image generation. We benchmarked our models on unguided, guided, and inpainting tasks using datasets like MNIST digits, Fashion MNIST, and CIFAR10, employing FID, SSIM, and PSNR metrics. We compared our models qualitatively to the quantum state-of-the-art, classical deep convolutional networks and U-Nets.

Our results show that our models vastly outperform the only other quantum denoising model by Dohun Kim \textit{et al.}~\cite{qddpm}. Additionally, our quantum models surpassed similarly-sized classical models and matched the efficacy of models twice their size. However, in inpainting tasks, classical models still hold an edge. We demonstrated the one-step generation capabilities of the first working unitary single-sampling model, both on quantum simulators and IBMQ hardware. 

In future studies, we aim to enhance variational quantum circuits by streamlining simulations using cached matrices, allowing for quicker GPU-parallel execution. Adopting 16-bit float precision could notably reduce RAM usage~\cite{low_precision_1,low_precision_2}, considering the demonstrated success in classical machine learning and prevalent GPU support for FP16. We're also keen to explore diffusion patching~\cite{diffpatch} which leverages pixel neighborhoods as channels, a method that might significantly boost execution speed, especially with RGB images. A deeper probe into optimal data embedding methods (\cref{sec:methods:guidance}) compared against classical models could yield insights into quantum knowledge representation. Furthermore, refining dense quantum circuits with a customized entangling circuit may offer superior spatial locality. Lastly, introducing classical components for post-processing in our models might present a pathway to circumvent quantum state normalization constraints and bolster overall performance.

%% file: content/X_supplements.tex
\clearpage
\setcounter{page}{1}
\maketitlesupplementary


\section{Preliminary Studies}
\label{sec:preliminary_studies}
In two preliminary studies, we examine the relationship between hyperparameters and sample quality metrics like FID, PSNR, and SSIM for the QU-Net architecture, as well as the impact of input scaling.

\begin{figure}[htb]
    \centering
    \subfloat
    [Q-U-Net Correlation]
    {
        \includegraphics[width=0.49\linewidth]{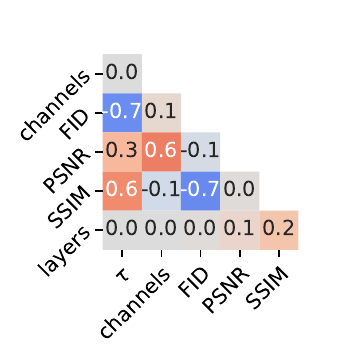}}
    \subfloat
    [U-Net Correlation]
    {
        \includegraphics[width=0.49\linewidth]{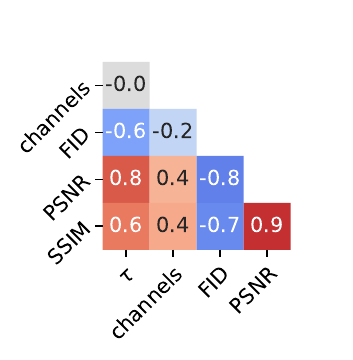}}
    \caption[Correlation of hyperparameters for QU-Nets and U-Nets]{
        Hyperparameter-sample quality correlation for QU-Nets and U-Nets on the Fashion dataset.
    }
    \label{fig:res:comparison:fashion:correlation}
\end{figure}

For quantum U-Nets, besides batch size and learning rate optimizations, we primarily consider the number of layers \(L\) and initial channels \(C\). Analysis reveals that a higher sampling step count \(\tau\) correlates with better sample quality. More channels show mixed results, and increasing the number of layers \(L\) marginally improves SSIM and PSNR.

Classic U-Nets, in contrast, display a strong correlation between channel number and sample quality across all metrics. From these findings, quantum U-Nets benefit from a larger \(\tau\), more channels for PSNR, and additional layers for SSIM.

Lastly, we analyze optimal input distributions for our unguided single-sample quantum models, computing them numerically as \(z = {U}^{-1} \cdot x\) due to the invertibility of the unitary diffusion matrix \(U\). This analysis shows that the real and imaginary parts of input vector \(z\) follow normal distributions \(\mathcal{N}(\mu=0.4, \sigma=0.24)\) and \(\mathcal{N}(\mu=0, \sigma=0.14)\) respectively. When manipulated by our diffusion model, they produce a non-uniform training data distribution \(x\) concentrated around 0 and 1 for dark and bright digit parts, respectively.

\begin{figure*}[bt]
    \centering
    \includegraphics[width=\linewidth]{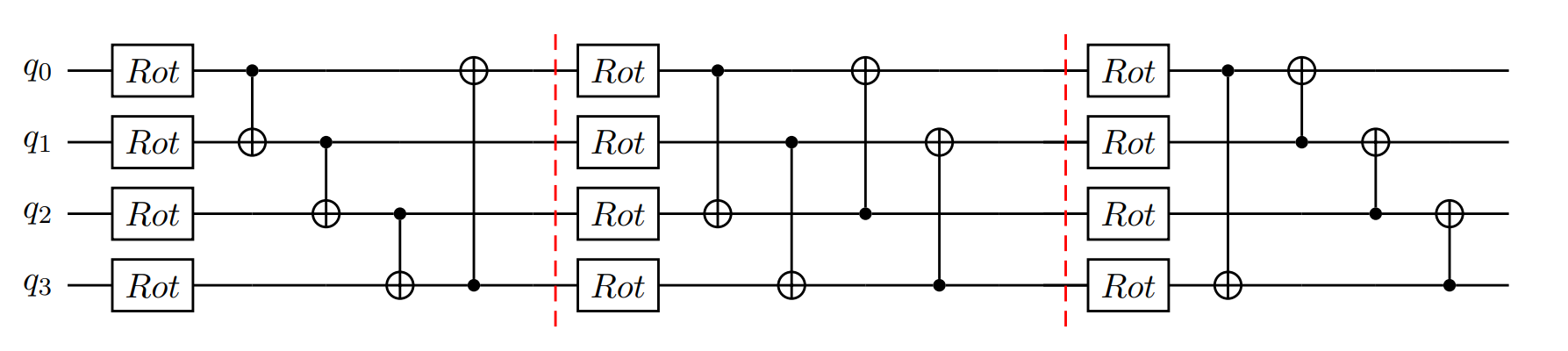}
    \caption{Example of the strongly entangling layers part of the VQC, where the red line denotes the layer boundary.}
    \label{fig:strongly}
\end{figure*}

\section{Quantum Architectures}
\begin{figure}[htb]
    \centering
    \subfloat[Unguided circuit]{
        \includegraphics[width=0.49\linewidth]{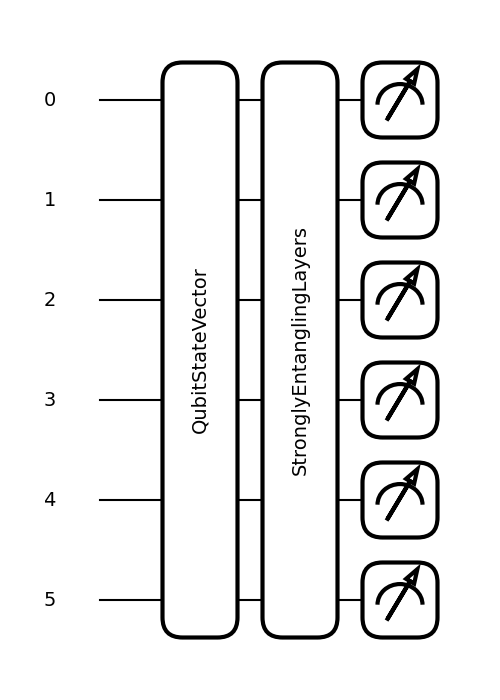}
        \label{fig:singlepassconstruct:a}%
    }
    \subfloat[Guided circuit]{
        \includegraphics[width=0.49\linewidth]{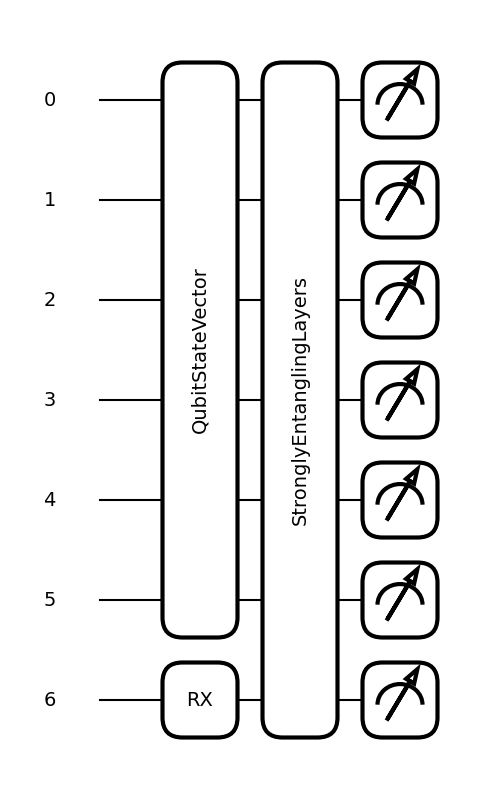}
        \label{fig:singlepassconstruct:c}%
    }  
    \caption[Single-Pass Models]{Unguided and guided single-sampling quantum circuits.}
    \label{fig:singlepassconstruct}
\end{figure}

In this work, we introduced two quantum architectures for denosing diffusion models. Each architecture has a guided and unguided variant as seen in \cref{fig:singlepassconstruct}. The guided variants all utilize a extra qubit called "ancilla", where the label is embedded. \cref{fig:singlepassconstruct:c} details how the label is embedded using a $R_X$ rotation (angle embedding) and how subsequent variational layers extend to the ancilla qubit. Both variants use strongly entangling layers as variational layers. In \cref{fig:strongly} we depicted a example of multiple strongly entangling layers, each trainable parameter \(\theta_i^j\) is associated with the \(i\)-th qubit and the \(j\)-th rotational gate, where \(j\) ranges over \(\{0,1,2\}\). For clarity, indices indicating the layer are omitted in the graphic representation. The entangling \text{CNOT} gates target the qubit at position \((i + l)\ \text{mod}\ n\), where \(l\) is the layer number. This scheme ensures circular entanglement in the first layer (\(l=1\)), with control and target qubits being adjacent, except for the last \text{CNOT} gate that completes the circle. In the second layer (\(l=2\)), the control and target qubits are separated by one qubit. For instance, the target qubit for the first qubit (\(i=0\)) in this layer is the third qubit (\(i=2\)).

\section{Hyperparameters}
\label{sec:hyperparameters}
Before we conducted our experiments, we ran a hyperparameter search for all tested model variants. We primarily focused our hyperparameter optimization on the learning rate, since it is often the most influencial hyperparameter. We used the bayesian optimization algorithm from the RayTune~\cite{raytune} library. All model variants with their respective hyperparameters are detailed in \cref{tab:lr-nq} for our classical models and \cref{tab:lr-q} for our quantum models.

\begin{table*}
    \centering
    \begin{tabular}{ |clllll| }
        \hline
        \textbf{Guided} & \textbf{Target} & \textbf{Dataset} & \textbf{Model}     & \textbf{Model shape}                              & \textbf{LR} \\
        \hline
        yes               & Data            & MNIST digits $8 \times 8$      & Deep Dense         & \twoline{$65$ and $64$ neurons,}{fully connected} & 0.0025      \\
                          &                 &                  & U-Net              & channel count $4, 8, 16$                          & 0.0057      \\
                          &                 &                  & U-Net              & channel count $2, 4, 8$                           & 0.0032      \\
                          &                 &                  & U-Net              & channel count $3, 6$                              & 0.0047      \\

        no                & Data            & MNIST digits $8 \times 8$      & Deep Convolutional & $1, 10, 10 1$                                     & 0.00759     \\
                          &                 &                  & Deep Convolutional & $1, 6, 6, 6, 6, 1$                                & 0.00013     \\
                          &                 &                  & U-Net              & channel count $2, 4, 8$                           & 0.00507     \\

        no                & Noise           & MNIST digits $8 \times 8$      & U-Net              & channel count $2, 4, 8$                           & 0.01339     \\
                          &                 &                  & Deep Convolutional & $1, 9, 10, 1$                                     & 0.00532     \\

        yes               & Noise           & MNIST $28 \times 28$           & U-Net              & channel count $3, 6$                              & 0.01719     \\
                          &                 &                  & U-Net              & channel count $2, 4, 8$                           & 0.00641     \\

        yes               & Noise           & Fashion MNIST         & Deep Convolutional & $1, 8, 8, 8, 8, 8, 8, 7, 1$                       & 0.01016     \\
                          &                 &                  & U-Net              & channel count $7, 14$                             & 0.01270     \\
        \hline
    \end{tabular}
    \caption[Learning rates of non-quantum models]{Learning rates of non-quantum models.
        The optimal learning rates have been chosen by a Bayesian optimization algorithm,
        facilitated by the \textit{RayTune}~\cite{raytune} library.
        MNIST digits $8 \times 8$ refer to the downsampled $8\times8$ images, MNIST digits $32 \times 32$ denotes upsampled images to $32\times32$.
        MNIST 28x28 refers to the original MNIST handwritten digits dataset~\cite{mnist},
        Fashion MNIST dataset~\cite{fashionmnist} uses $32\times32$ images.
    }
    \label{tab:lr-nq}
\end{table*}

\begin{table*}
    \centering
    \begin{tabular}{ |lclllll| }
        \hline
         & \textbf{Guided} & \textbf{Target} & \textbf{Dataset} & \textbf{Model} & \textbf{Model shape}                               & \textbf{LR} \\
        \hline
         & yes               & Data            & MNIST digits $8 \times 8$      & Q-Dense        & 47 layers                                          & 0.00097     \\
         &                   &                 &                  & Q-Dense Re-Up  & 47 layers, 3 re-ups                                & 0.00360     \\
         &                   &                 &                  & Q-Dense Re-Up  & 47 layers, 5 re-ups                                & 0.00345     \\
         &                   &                 &                  & Q-Dense Re-Up  & 47 layers, 7 re-ups                                & 0.00362     \\

         & no                & Data            & MNIST digits $8 \times 8$      & Q-Dense        & 50 layers                                          & 0.00065     \\
         &                   &                 &                  & Q-U-Net Simple & \twoline{Channel count $2, 4, 8$}{8 Layers each}   & 0.00023     \\
         &                   &                 &                  & Q-U-Net Simple & \twoline{Channel count $2, 4, 8$}{12 Layers each}  & 0.00815     \\

         & no                & Noise           & MNIST digits $8 \times 8$      & Q-Dense        & 55 layers                                          & 0.00160     \\
         &                   &                 &                  & Q-U-Net Simple & \twoline{Channel count $2, 4, 8$}{8 Layers each}   & 0.00113     \\
         &                   &                 &                  & Q-U-Net Simple & \twoline{Channel count $2, 4, 8$}{12 Layers each}  & 0.00912     \\

         & yes               & Noise           & MNIST digits $28 \times 28$           & Q-Dense        & 30 layers                                          & 0.00409     \\
         &                   &                 &                  & Q-Dense        & 60 layers                                          & 0.00211     \\
         &                   &                 &                  & Q-U-Net Simple & \twoline{Channel count $2, 4, 8$}{9 Layers each}   & 0.00287     \\
         &                   &                 &                  & Q-U-Net Simple & \twoline{Channel count $2, 4, 8$}{19 Layers each}  & 0.01479     \\
         & yes               & Noise           & Fashion          & Q-Dense        & 121 layers                                         & 0.00014     \\
         &                   &                 &                  & Q-Dense        & 60 layers                                          & 0.00723     \\
         &                   &                 &                  & Q-U-Net Simple & \twoline{Channel count $3, 6, 12$}{8 Layers each}  & 0.00051     \\
         &                   &                 &                  & Q-U-Net Simple & \twoline{Channel count $3, 6, 12$}{12 Layers each} & 0.00029     \\

         & no                & Data            & MNIST digits $8 \times 8$      & Single Sample  & 476 Layers, with ancilla                           & 0.00012     \\
         & no                &                 & MNIST digits $8 \times 8$      & Single Sample  & 47 Layers, with ancilla                            & 0.00322     \\
         & no                &                 & MNIST digits $8 \times 8$      & Single Sample  & 55 Layers, no ancilla                              & 0.00172     \\
         & no                &                 & MNIST digits $8 \times 8$      & Single Sample  & 555 Layers, no ancilla                             & 0.00002     \\
         & no                &                 & MNIST digits $8 \times 8$      & Single Sample  & 111 Layers, no ancilla                             & 0.00419     \\
         & no                &                 & MNIST digits $8 \times 8$      & Single Sample  & 95 Layers, with ancilla                            & 0.00104     \\
         & yes               &                 & MNIST digits $8 \times 8$      & Single Sample  & 476 Layers                                         & 0.00022     \\
         & yes               &                 & MNIST digits $8 \times 8$      & Single Sample  & 476 layers, 10 re-ups                              & 0.00016     \\
         & yes               &                 & MNIST digits $8 \times 8$      & Single Sample  & 47 Layers                                          & 0.00721     \\
         & yes               &                 & MNIST digits $8 \times 8$      & Single Sample  & 47 layers, 10 re-ups                               & 0.00220     \\
         & yes               &                 & MNIST digits $8 \times 8$      & Single Sample  & 95 Layers                                          & 0.00099     \\
         & no                &                 & MNIST digits $32 \times 32$        & Single Sample  & 66 Layers, no ancilla                              & 0.00212     \\
         & no                &                 & MNIST digits $32 \times 32$        & Single Sample  & 133 Layers, no ancilla                             & 0.00190     \\
        \hline
    \end{tabular}
    \caption[Learning rates of quantum models]{Learning rates of quantum models.
        If not further specified, quantum layers are strongly entangling layers as described in~\cite{ccqc}.
        The datasets are the same as in \cref{tab:lr-nq}.
        All models used a batch size of 20 (except the quantum U-Nets, which used a batch size of 10),
        as higher batch sizes, while slightly improving the progress per iteration,
        lead to a higher number of crashes due to insufficient memory.
    }
    \label{tab:lr-q}
\end{table*}

%% file: main.bbl
\begin{thebibliography}{55}
\providecommand{\natexlab}[1]{#1}
\providecommand{\url}[1]{\texttt{#1}}
\expandafter\ifx\csname urlstyle\endcsname\relax
  \providecommand{\doi}[1]{doi: #1}\else
  \providecommand{\doi}{doi: \begingroup \urlstyle{rm}\Url}\fi

\bibitem[Barenco et~al.(1995)Barenco, Bennett, Cleve, DiVincenzo, Margolus, Shor, Sleator, Smolin, and Weinfurter]{q_ops}
Adriano Barenco, Charles~H. Bennett, Richard Cleve, David~P. DiVincenzo, Norman Margolus, Peter Shor, Tycho Sleator, John~A. Smolin, and Harald Weinfurter.
\newblock Elementary gates for quantum computation.
\newblock \emph{Physical Review A}, 52\penalty0 (5):\penalty0 3457--3467, 1995.

\bibitem[Bergholm et~al.(2022)Bergholm, Izaac, Schuld, Gogolin, Ahmed, Ajith, Alam, Alonso-Linaje, AkashNarayanan, and Asadi]{pennylane}
Ville Bergholm, Josh Izaac, Maria Schuld, Christian Gogolin, Shahnawaz Ahmed, Vishnu Ajith, M.~Sohaib Alam, Guillermo Alonso-Linaje, B. AkashNarayanan, and Ali Asadi.
\newblock {PennyLane}: Automatic differentiation of hybrid quantum-classical computations, 2022.

\bibitem[Biamonte et~al.(2017)Biamonte, Wittek, Pancotti, Rebentrost, Wiebe, and Lloyd]{qml1}
Jacob Biamonte, Peter Wittek, Nicola Pancotti, Patrick Rebentrost, Nathan Wiebe, and Seth Lloyd.
\newblock Quantum machine learning.
\newblock \emph{Nature}, 549\penalty0 (7671):\penalty0 195--202, 2017.

\bibitem[Cong et~al.(2019)Cong, Choi, and Lukin]{qconv2}
Iris Cong, Soonwon Choi, and Mikhail~D. Lukin.
\newblock Quantum convolutional neural networks.
\newblock \emph{Nature Physics}, 15\penalty0 (12):\penalty0 1273--1278, 2019.

\bibitem[Cramer et~al.(2010)Cramer, Plenio, Flammia, Somma, Gross, Bartlett, Landon-Cardinal, Poulin, and Liu]{cramer2010efficient}
Marcus Cramer, Martin~B Plenio, Steven~T Flammia, Rolando Somma, David Gross, Stephen~D Bartlett, Olivier Landon-Cardinal, David Poulin, and Yi-Kai Liu.
\newblock Efficient quantum state tomography.
\newblock \emph{Nature communications}, 1\penalty0 (1):\penalty0 149, 2010.

\bibitem[Deng(2012)]{mnist}
Li Deng.
\newblock The {MNIST} database of handwritten digit images for machine learning research.
\newblock \emph{IEEE Signal Processing Magazine}, 29\penalty0 (6):\penalty0 141--142, 2012.

\bibitem[Dhariwal and Nichol(2021)]{diffusion_beats_gan}
Prafulla Dhariwal and Alex Nichol.
\newblock Diffusion models beat {GANs} on image synthesis, 2021.

\bibitem[Drozdzal et~al.(2016)Drozdzal, Vorontsov, Chartrand, Kadoury, and Pal]{skip2}
Michal Drozdzal, Eugene Vorontsov, Gabriel Chartrand, Samuel Kadoury, and Chris Pal.
\newblock The importance of skip connections in biomedical image segmentation, 2016.

\bibitem[Elasri et~al.(2022)Elasri, Elharrouss, Al-ma'adeed, and Tairi]{imagegeneration1}
Mohamed Elasri, Omar Elharrouss, Somaya Al-ma'adeed, and Hamid Tairi.
\newblock Image generation: A review.
\newblock \emph{Neural Processing Letters}, 54, 2022.

\bibitem[Gabor et~al.(2020)Gabor, Sünkel, Ritz, Phan, Belzner, Roch, Feld, and Linnhoff-Popien]{qcaccel}
Thomas Gabor, Leo Sünkel, Fabian Ritz, Thomy Phan, Lenz Belzner, Christoph Roch, Sebastian Feld, and Claudia Linnhoff-Popien.
\newblock The holy grail of quantum artificial intelligence: Major challenges in accelerating the machine learning pipeline, 2020.

\bibitem[Goodfellow et~al.(2020)Goodfellow, Pouget-Abadie, Mirza, Xu, Warde-Farley, Ozair, Courville, and Bengio]{gan_goodfellow}
Ian Goodfellow, Jean Pouget-Abadie, Mehdi Mirza, Bing Xu, David Warde-Farley, Sherjil Ozair, Aaron Courville, and Yoshua Bengio.
\newblock Generative adversarial networks.
\newblock \emph{Communications of the ACM}, 63\penalty0 (11):\penalty0 139--144, 2020.

\bibitem[Gu et~al.(2017)Gu, Wang, Kuen, Ma, Shahroudy, Shuai, Liu, Wang, Wang, Wang, Cai, and Chen]{cnn3}
Jiuxiang Gu, Zhenhua Wang, Jason Kuen, Lianyang Ma, Amir Shahroudy, Bing Shuai, Ting Liu, Xingxing Wang, Li Wang, Gang Wang, Jianfei Cai, and Tsuhan Chen.
\newblock Recent advances in convolutional neural networks, 2017.

\bibitem[Guan et~al.(2019)Guan, Khan, Sikdar, and Chitnis]{unet2}
Steven Guan, Amir~A Khan, Siddhartha Sikdar, and Parag~V Chitnis.
\newblock Fully dense {U-Net} for 2-d sparse photoacoustic tomography artifact removal.
\newblock \emph{IEEE journal of biomedical and health informatics}, 24\penalty0 (2):\penalty0 568--576, 2019.

\bibitem[Gupta et~al.(2015)Gupta, Agrawal, Gopalakrishnan, and Narayanan]{low_precision_2}
Suyog Gupta, Ankur Agrawal, Kailash Gopalakrishnan, and Pritish Narayanan.
\newblock Deep learning with limited numerical precision, 2015.

\bibitem[Henderson et~al.(2019)Henderson, Shakya, Pradhan, and Cook]{qconv1}
Maxwell Henderson, Samriddhi Shakya, Shashindra Pradhan, and Tristan Cook.
\newblock Quanvolutional neural networks: Powering image recognition with quantum circuits, 2019.

\bibitem[Heusel et~al.(2018)Heusel, Ramsauer, Unterthiner, Nessler, and Hochreiter]{fid1}
Martin Heusel, Hubert Ramsauer, Thomas Unterthiner, Bernhard Nessler, and Sepp Hochreiter.
\newblock {GANs} trained by a two time-scale update rule converge to a local nash equilibrium, 2018.

\bibitem[Ho et~al.(2020)Ho, Jain, and Abbeel]{ddpm}
Jonathan Ho, Ajay Jain, and Pieter Abbeel.
\newblock Denoising diffusion probabilistic models.
\newblock In \emph{Advances in Neural Information Processing Systems}, pages 6840--6851. Curran Associates, Inc., 2020.

\bibitem[Huang et~al.(2018)Huang, Yu, and Wang]{imagegeneration2}
He Huang, Philip~S. Yu, and Changhu Wang.
\newblock An introduction to image synthesis with generative adversarial nets, 2018.

\bibitem[Huang et~al.(2020)Huang, Lin, Tong, Hu, Zhang, Iwamoto, Han, Chen, and Wu]{unet3}
Huimin Huang, Lanfen Lin, Ruofeng Tong, Hongjie Hu, Qiaowei Zhang, Yutaro Iwamoto, Xianhua Han, Yen-Wei Chen, and Jian Wu.
\newblock {U-Net} 3+: A full-scale connected {U-Net} for medical image segmentation.
\newblock In \emph{ICASSP 2020-2020 IEEE International Conference on Acoustics, Speech and Signal Processing (ICASSP)}, pages 1055--1059. IEEE, 2020.

\bibitem[Kim and Kang(2023)]{qddpm}
Dohun Kim and Seokhyeong Kang.
\newblock Quantum denoising diffusion probabilistic models for image generation.
\newblock In \emph{Korean Conference on Semiconductors}, 2023.

\bibitem[Kingma and Ba(2017)]{adam}
Diederik~P. Kingma and Jimmy Ba.
\newblock Adam: A method for stochastic optimization, 2017.

\bibitem[Krizhevsky(2012)]{cifar}
Alex Krizhevsky.
\newblock Learning multiple layers of features from tiny images.
\newblock \emph{University of Toronto}, 2012.

\bibitem[Krizhevsky et~al.(2017)Krizhevsky, Sutskever, and Hinton]{cnn1}
Alex Krizhevsky, Ilya Sutskever, and Geoffrey~E Hinton.
\newblock Imagenet classification with deep convolutional neural networks.
\newblock \emph{Communications of the ACM}, 60\penalty0 (6):\penalty0 84--90, 2017.

\bibitem[Kölle et~al.(2022)Kölle, Giovagnoli, Stein, Mansky, Hager, and Linnhoff-Popien]{weightremapping}
Michael Kölle, Alessandro Giovagnoli, Jonas Stein, Maximilian~Balthasar Mansky, Julian Hager, and Claudia Linnhoff-Popien.
\newblock Improving convergence for quantum variational classifiers using weight re-mapping, 2022.

\bibitem[LeCun et~al.(2010)LeCun, Kavukcuoglu, and Farabet]{cnn2}
Yann LeCun, Koray Kavukcuoglu, and Cl{\'e}ment Farabet.
\newblock Convolutional networks and applications in vision.
\newblock In \emph{Proceedings of 2010 IEEE international symposium on circuits and systems}, pages 253--256. IEEE, 2010.

\bibitem[Liaw et~al.(2018)Liaw, Liang, Nishihara, Moritz, Gonzalez, and Stoica]{raytune}
Richard Liaw, Eric Liang, Robert Nishihara, Philipp Moritz, Joseph~E Gonzalez, and Ion Stoica.
\newblock Tune: A research platform for distributed model selection and training.
\newblock \emph{arXiv:1807.05118}, 2018.

\bibitem[Liu et~al.(2018)Liu, Wei, Lu, and Zhou]{fid2}
Shaohui Liu, Yi Wei, Jiwen Lu, and Jie Zhou.
\newblock An improved evaluation framework for generative adversarial networks, 2018.

\bibitem[Lloyd et~al.(2013)Lloyd, Mohseni, and Rebentrost]{qml2}
Seth Lloyd, Masoud Mohseni, and Patrick Rebentrost.
\newblock Quantum algorithms for supervised and unsupervised machine learning, 2013.

\bibitem[Lugmayr et~al.(2022)Lugmayr, Danelljan, Romero, Yu, Timofte, and Gool]{inpainting}
Andreas Lugmayr, Martin Danelljan, Andr{\'{e}}s Romero, Fisher Yu, Radu Timofte, and Luc~Van Gool.
\newblock Repaint: Inpainting using denoising diffusion probabilistic models.
\newblock \emph{CoRR}, abs/2201.09865, 2022.

\bibitem[Luhman and Luhman(2022)]{diffpatch}
Troy Luhman and Eric Luhman.
\newblock Improving diffusion model efficiency through patching, 2022.

\bibitem[Mitarai et~al.(2018)Mitarai, Negoro, Kitagawa, and Fujii]{acl_mitarai}
K. Mitarai, M. Negoro, M. Kitagawa, and K. Fujii.
\newblock Quantum circuit learning.
\newblock \emph{Physical Review A}, 98\penalty0 (3), 2018.

\bibitem[Möttönen et~al.(2004)Möttönen, Vartiainen, Bergholm, and Salomaa]{mottonen}
Mikko Möttönen, Juha~J. Vartiainen, Ville Bergholm, and Martti~M. Salomaa.
\newblock Transformation of quantum states using uniformly controlled rotations, 2004.

\bibitem[Nichol and Dhariwal(2021)]{ddpm2}
Alex Nichol and Prafulla Dhariwal.
\newblock Improved denoising diffusion probabilistic models, 2021.

\bibitem[Nichol et~al.(2022)Nichol, Dhariwal, Ramesh, Shyam, Mishkin, McGrew, Sutskever, and Chen]{glide}
Alex Nichol, Prafulla Dhariwal, Aditya Ramesh, Pranav Shyam, Pamela Mishkin, Bob McGrew, Ilya Sutskever, and Mark Chen.
\newblock {GLIDE}: Towards photorealistic image generation and editing with text-guided diffusion models, 2022.

\bibitem[Oh et~al.(2020)Oh, Choi, and Kim]{qconv3}
Seunghyeok Oh, Jaeho Choi, and Joongheon Kim.
\newblock A tutorial on quantum convolutional neural networks ({QCNN}), 2020.

\bibitem[OpenAI(2023)]{gpt4}
OpenAI.
\newblock Gpt-4 technical report, 2023.

\bibitem[Pathak et~al.(2016)Pathak, Kr{\"{a}}henb{\"{u}}hl, Donahue, Darrell, and Efros]{inpainting2}
Deepak Pathak, Philipp Kr{\"{a}}henb{\"{u}}hl, Jeff Donahue, Trevor Darrell, and Alexei~A. Efros.
\newblock Context encoders: Feature learning by inpainting.
\newblock \emph{CoRR}, abs/1604.07379, 2016.

\bibitem[P{\'{e}}rez-Salinas et~al.(2020)P{\'{e}}rez-Salinas, Cervera-Lierta, Gil-Fuster, and Latorre]{datareuploading}
Adri{\'{a}}n P{\'{e}}rez-Salinas, Alba Cervera-Lierta, Elies Gil-Fuster, and Jos{\'{e}}~I. Latorre.
\newblock Data re-uploading for a universal quantum classifier.
\newblock \emph{Quantum}, 4:\penalty0 226, 2020.

\bibitem[Preskill(2018)]{nisq}
John Preskill.
\newblock Quantum computing in the {NISQ} era and beyond.
\newblock \emph{Quantum}, 2:\penalty0 79, 2018.

\bibitem[Rombach et~al.(2021)Rombach, Blattmann, Lorenz, Esser, and Ommer]{stablediffusion}
Robin Rombach, Andreas Blattmann, Dominik Lorenz, Patrick Esser, and Björn Ommer.
\newblock High-resolution image synthesis with latent diffusion models, 2021.

\bibitem[Ronneberger et~al.(2015)Ronneberger, Fischer, and Brox]{unet}
Olaf Ronneberger, Philipp Fischer, and Thomas Brox.
\newblock {U-Net}: Convolutional networks for biomedical image segmentation, 2015.

\bibitem[Schuld et~al.(2019)Schuld, Bergholm, Gogolin, Izaac, and Killoran]{qgrad}
Maria Schuld, Ville Bergholm, Christian Gogolin, Josh Izaac, and Nathan Killoran.
\newblock Evaluating analytic gradients on quantum hardware.
\newblock \emph{Physical Review A}, 99\penalty0 (3), 2019.

\bibitem[Schuld et~al.(2020)Schuld, Bocharov, Svore, and Wiebe]{ccqc}
Maria Schuld, Alex Bocharov, Krysta~M. Svore, and Nathan Wiebe.
\newblock Circuit-centric quantum classifiers.
\newblock \emph{Physical Review A}, 101\penalty0 (3), 2020.

\bibitem[Sohl-Dickstein et~al.(2015)Sohl-Dickstein, Weiss, Maheswaranathan, and Ganguli]{dpm_sohl}
Jascha Sohl-Dickstein, Eric Weiss, Niru Maheswaranathan, and Surya Ganguli.
\newblock Deep unsupervised learning using nonequilibrium thermodynamics.
\newblock In \emph{International Conference on Machine Learning}, pages 2256--2265. PMLR, 2015.

\bibitem[Song et~al.(2020)Song, Meng, and Ermon]{ddim}
Jiaming Song, Chenlin Meng, and Stefano Ermon.
\newblock Denoising diffusion implicit models, 2020.

\bibitem[Song et~al.(2023)Song, Dhariwal, Chen, and Sutskever]{consistency}
Yang Song, Prafulla Dhariwal, Mark Chen, and Ilya Sutskever.
\newblock Consistency models, 2023.

\bibitem[Steane(1998)]{qc_steane}
Andrew Steane.
\newblock Quantum computing.
\newblock \emph{Reports on Progress in Physics}, 61\penalty0 (2):\penalty0 117--173, 1998.

\bibitem[Szegedy et~al.(2015)Szegedy, Liu, Jia, Sermanet, Reed, Anguelov, Erhan, Vanhoucke, and Rabinovich]{inceptionv3}
Christian Szegedy, Wei Liu, Yangqing Jia, Pierre Sermanet, Scott Reed, Dragomir Anguelov, Dumitru Erhan, Vincent Vanhoucke, and Andrew Rabinovich.
\newblock Going deeper with convolutions.
\newblock In \emph{Proceedings of the IEEE conference on computer vision and pattern recognition}, pages 1--9, 2015.

\bibitem[Tong et~al.(2017)Tong, Li, Liu, and Gao]{skip}
Tong Tong, Gen Li, Xiejie Liu, and Qinquan Gao.
\newblock Image super-resolution using dense skip connections.
\newblock In \emph{Proceedings of the IEEE international conference on computer vision}, pages 4799--4807, 2017.

\bibitem[Vaswani et~al.(2017)Vaswani, Shazeer, Parmar, Uszkoreit, Jones, Gomez, Kaiser, and Polosukhin]{attention}
Ashish Vaswani, Noam Shazeer, Niki Parmar, Jakob Uszkoreit, Llion Jones, Aidan~N. Gomez, Lukasz Kaiser, and Illia Polosukhin.
\newblock Attention is all you need, 2017.

\bibitem[Venkatesh et~al.(2016)Venkatesh, Nurvitadhi, and Marr]{low_precision_1}
Ganesh Venkatesh, Eriko Nurvitadhi, and Debbie Marr.
\newblock Accelerating deep convolutional networks using low-precision and sparsity, 2016.

\bibitem[Verdon et~al.(2019)Verdon, Broughton, and Biamonte]{qml3}
Guillaume Verdon, Michael Broughton, and Jacob Biamonte.
\newblock A quantum algorithm to train neural networks using low-depth circuits, 2019.

\bibitem[Wang et~al.(2003)Wang, Simoncelli, and Bovik]{ssim}
Z. Wang, E.P. Simoncelli, and A.C. Bovik.
\newblock Multiscale structural similarity for image quality assessment.
\newblock In \emph{The Thrity-Seventh Asilomar Conference on Signals, Systems and Computers, 2003}, pages 1398--1402 Vol.2, 2003.

\bibitem[Xiao et~al.(2017)Xiao, Rasul, and Vollgraf]{fashionmnist}
Han Xiao, Kashif Rasul, and Roland Vollgraf.
\newblock {Fashion-MNIST}: a novel image dataset for benchmarking machine learning algorithms, 2017.

\bibitem[Zhang et~al.(2022)Zhang, Li, and Yuan]{PhysRevLett.129.230504}
Xiao-Ming Zhang, Tongyang Li, and Xiao Yuan.
\newblock Quantum state preparation with optimal circuit depth: Implementations and applications.
\newblock \emph{Phys. Rev. Lett.}, 129:\penalty0 230504, 2022.

\end{thebibliography}
